\documentclass[twocolumn,pre,a4paper,showpacs,floatfix]{revtex4}
\usepackage{amsmath}
\usepackage{graphicx}

\newcommand{\df}{{d_{\!f}}}
\newcommand{\letter}{note}
\newcommand{\latin}[1]{\emph{#1}}
\newcommand{\ie}{\latin{i.$\,$e.}}
\newcommand{\aposteriori}{\latin{a posteriori}}
\newcommand{\insitu}{\latin{in situ}}
\newcommand{\etal}{\latin{et al.}}

\newcommand{\german}[1]{\emph{#1}}
\newcommand{\ansatz}{\german{ansatz}}

\newcommand{\tgel}{\tau_{\text{gel}}}
\newcommand{\tppd}{\tau_{\text{ppd}}}
\newcommand{\tcld}{\tau_{\text{cld}}}
\newcommand{\kagg}{k_{\text{agg}}}
\newcommand{\kdep}{k_{\text{dep}}}

\begin{document}

\title{Scaling theory for diffusion limited cluster aggregation in a
  porous medium}

\author{Patrick B. Warren}
\affiliation{Unilever R\&D Port Sunlight, Bebington, Wirral, CH63 3JW, UK.}

\date{\today}

\begin{abstract}
A scaling theory is developed for diffusion-limited cluster
aggregation in a porous medium, where the primary particles and
clusters stick irreversibly to the walls of the pore space as well as
to each other.  Three scaling regimes are predicted, connected by
smooth crossovers.  The first regime is at low primary particle
concentrations where the primary particles stick individually to the
walls.  The second regime is at intermediate concentrations where
clusters grow to a certain size, smaller than the pore size, then
stick individually to the walls.  The third regime is at high
concentrations where the final state is a pore-space-filling network.
\end{abstract}



\maketitle

Diffusion-limited aggregation (DLA) and diffusion-limited cluster
aggregation (DLCA) attracted much attention from the early 1980s
following the discovery of scale-invariant (fractal) structures
\cite{WS81, DLCA83, WO84}.  In DLA, primary particles undergo random
walks before adhering irreversibly to a growing cluster \cite{WS81}.
In DLCA, both primary particles and growing clusters undergo random
walks, adhering to each other when they come into contact
\cite{DLCA83}.  Despite a couple of decades or so of intense study,
there still remain interesting questions about these processes.  On
the one hand deep puzzles linger about the origins of the scale
invariance and whether a simple fractal picture suffices \cite{DLA08},
and on the other hand generalisations to more complicated situations
have been considered, such as DLCA as a function of primary particle
concentration \cite{DLCAconc} or in the presence of sedimentation
\cite{DLCAsed} or flow \cite{PU90, WB92}.

In this \letter\ I present a scaling analysis of DLCA as a function of
the primary particle concentration for the case where the primary
particles and clusters can adhere irreversibly to the walls of the
porous medium as well as to each other.  I consider the case where the
primary particles are much smaller than the pores.  My motivation is
to gain a deeper understanding of the deposition of small particles
into porous materials, with applications that range from deep-bed
filtration \cite{deepbed}, to the aggregation and deposition of
colloidal asphaltene in reservoir rocks in oil recovery
\cite{XBLC+08}.  The results presented here are also hopefully
intrinsically interesting, although the model still lacks certain
features present in real systems such as flow.  This will be discussed
further at the end.  Previously, DLA has been studied in confined
geometries \cite{DLAwedge}, but this is not really relevant to the
present problem.

Let me start by summarising what is known about DLCA \cite{DLCAbooks}.
Firstly, as already mentioned, the clusters have a scale invariant
structure such that the aggregation number $N$ of a cluster of linear
size $R$ obeys
\begin{equation}
N\sim (R/a_0)^{\df}\label{eq:df}
\end{equation}
where $a_0$ is the primary particle size and $\df$ is the fractal
dimension.  This scaling law has been confirmed both in experiments on
aggregating colloids and in computer simulations \cite{WO84, Mea84}.
For the remainder of this \letter\ my focus will be on the practically
relevant case of three dimensions, where $\df\approx 1.75$ \cite{WO84, Mea84}.

A second line of enquiry into DLCA has been into the kinetics of the
process \cite{LeyvDE, XWLH+85, Meakin92}.  A long time ago,
Smoluchowski presented a mean-field theory to describe the evolution
of the cluster size distribution \cite{Smol16}.  When updated to take
account of the fractal structure of the aggregates, Smoluchowski's
theory has turned out to be suprisingly accurate.  One result of
Smoluchowski's theory is that the cluster size distribution remains
\emph{bell-shaped} \cite{LeyvDE, XWLH+85}.  In scaling terms this
justifies the notion of a \emph{characteristic cluster size}.  Armed
with this foreknowledge, it is possible to use a simple scaling
argument to determine how the characteristic cluster size grows with
time \cite{Meakin92}.  In fact this is the key to the development,
since the scaling argument generalises in a straightforward way to the
problem of DLCA in a porous medium, by taking account of the depletion
kinetics of clusters as they adhere to the walls.

\begin{figure}
\begin{center}
\includegraphics{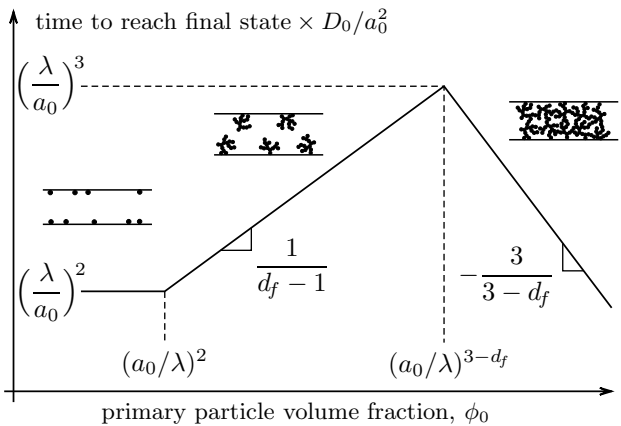}
\end{center}
\caption{Double-logarithmic plot showing scaling regimes for
  diffusion-limited cluster aggregation in a porous medium, here
  represented schematically as a slit.  In order of increasing volume
  fraction, the regimes are primary particle deposition, cluster
  deposition, and formation of a pore-space-filling network.
  Parameters are the primary particle size $a_0$, the characteristic
  pore size $\lambda\gg a_0$, the primary particle diffusion
  coefficient $D_0$, and the fractal dimension $\df$.  In three
  dimensions $\df\approx1.75$ and the slopes of the lines are
  $1/(\df-1)\approx1.33$ and
  $-3/(3-\df)\approx-2.40$.\label{fig:1}}
\end{figure}

The kinetic scaling argument for pure DLCA runs as follows.  Let the
$n(t)$ be the cluster number density at time $t$.  Then, in mean field
theory, $dn/dt=-\kagg n$ where $\kagg$ is the aggregation rate for the
characteristic clusters.  Again taking a mean field approach, this
aggregation rate is given by the classic Smoluchowski
diffusion-limited reaction rate $\kagg\sim RDn$, where $R$ is the
characteristic cluster size and $D$ is the corresponding diffusion
coefficient.  This result can be derived by considering the steady
state diffusive flux to a sphere of radius $\sim R$ from a
concentration field of number density $n$ at infinity.  Now, the
Stokes-Einstein relation implies $D\sim (a_0/R)\times D_0$ where $D_0$
is the primary particle diffusion coefficient.  One concludes that
$dn/dt\sim -a_0D_0n^2$.  Integrating this gives $1/n-1/n_0\sim a_0D_0
t$, where $n_0$ is the number density of primary particles.  Defining
the aggregation number for the characteristic clusters to be $N\equiv
n_0/n$, one finds
\begin{equation}
N=1+k_0t\label{eq:smol}
\end{equation}
where $k_0\sim a_0D_0n_0$ is the characteristic aggregation time for
primary particles.

Remarkably, this is exactly the same answer as can be obtained from
the full Smoluchowski theory.  For $k_0t\gg1$ it implies $N\sim t^z$
with $z=1$ being the kinetic scaling exponent.  This has been
confirmed both in experiment and in computer simulations \cite{XWLH+85,
  Meakin92}.

The fractal nature of the growing clusters means that they eventually
expand to fill space, forming a network or gelled state.  The time
scale for this can be worked out by considering the effective cluster
volume fraction $\phi\sim nR^3$.  Assuming $k_0t\gg1$ and using
Eqs.~\eqref{eq:df} and \eqref{eq:smol}, one finds
$\phi\sim\phi_0(k_0t)^{3/\df-1}$ where $\phi_0\sim na_0^3$ is the
volume fraction of primary particles.  Since $\df<3$, $\phi$ increases
with time and gelation occurs when $\phi\to1$.  The gelation time is
therefore
\begin{equation}
\tgel\sim a_0^2/D_0\times\phi_0^{-3/(3-\df)}\,.\label{eq:tgel}
\end{equation}
A point to note in idealised DLCA is that the final state of the
system is \emph{always} a gelled state.  Experimentally other factors may
intervene of course, such as sedimentation or creaming, or cluster
rupture due to flow.

The framework has now been set for the generalisation to DLCA in a
porous medium.  In this more complex situation, the basic idea is that
there is a competition between the rate at which primary particles and
clusters find each other, and the rate at which they are depleted by
becoming stuck to the walls.  I consider the case where the porous
medium has a unimodal pore size distribution and therefore a
characteristic pore size $\lambda\gg a_0$ (the generalisation to
multiple length scales or fractal pore spaces is not too difficult).
I also take the starting point to be a random dispersion of primary
particles at a number density $n_0$ (volume fraction $\phi_0\sim
n_0a_0^3$).  It might be objected that this is experimentally
unfeasible but this point will be discussed in more depth at the end.

\begin{figure}
\begin{center}
\includegraphics{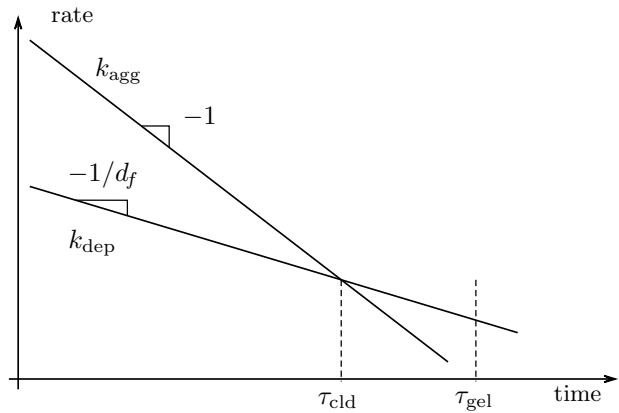}
\end{center}
\caption{Intermediate regime of Fig.~\ref{fig:1}:
  double-logarithmic plot showing how the cluster-cluster aggregation
  rate falls faster than the cluster-wall depletion rate as a function
  of time.  Where the two rates cross over, clusters become attached
  to the walls, halting the aggregation process.  The time at this
  cross over point should be less than the gelation time, otherwise
  wall cluster deposition is pre-empted by gelation.  In three
  dimension, $\df\approx1.75$ and so
  $1/\df\approx0.57$.\label{fig:2}}
\end{figure}

Since $\lambda^2/D$ is the time it takes for the cluster of size $R$
to diffuse a distance of order the pore size, the rate at which such
clusters are depleted by sticking to the walls is given in scaling
terms by $\kdep\sim D/\lambda^2$.  This assumes $R\ll\lambda$, which
will be verified \aposteriori.  This rate should be compared to the
aggregation rate from Smoluchowski's theory above.  Both are changing
in time and to compare on a like-for-like basis the $t$-dependence
should be explicitly extracted.  One finds
\begin{equation}
\kagg\sim\frac{k_0}{1+k_0t},\qquad
\kdep\sim\frac{D_0}{\lambda^2}
(1+k_0t)^{-1/\df}.\label{eq:rates}
\end{equation}

Two situations should be considered, depending on the initial values
of the rates for primary particles.  Firstly, if $\kagg\ll\kdep$ at
$t=0$, then the primary particles diffuse to the walls much
faster than they undergo aggregation.  This condition implies
$k_0\ll{D_0}/{\lambda^2}$ or
\begin{equation}
\phi_0\ll({a_0}/{\lambda})^2.
\label{eq:ppd}
\end{equation}
In this limit aggregation is irrelevant and the final state is one
where primary particles are stuck to the walls.  Moreover, since there
are $n_0$ particles per unit volume spread over the walls at a surface
area per unit volume $\sim1/\lambda$, the area fraction covered by
particles is $a_0^2n_0\lambda\sim\phi_0\lambda/a_0$.  But, using
Eq.~\eqref{eq:ppd}, $\phi_0\lambda/a_0\ll a_0/\lambda\ll1$.  In other
words the primary particles are very sparsely distributed over the
walls.  Finally the time to reach the final state is
$\tppd\sim\lambda^2/D_0$ where $D_0$ is the primary particle diffusion
coefficient.  This can be written as
\begin{equation}
\tppd\sim a_0^2/D_0\times({\lambda}/{a_0})^2.\label{eq:tppd}
\end{equation}  
This is independent of the primary particle concentration of course.

Now consider the case where $\kagg\gg\kdep$ at $t=0$.  In this
situation, the primary particles start to aggregate before they see
the walls.  Eq.~\eqref{eq:rates} shows that, as time progresses,
$\kagg$ falls faster than $\kdep$, since $\df>1$.  Eventually the wall
depletion rate must overtake the aggregation rate.  This is strongly
suggestive of a crossover, in other words one expects aggregation to
proceed as in standard DLCA until $\kagg\sim\kdep$, at which point the
clusters are removed by sticking to the walls.  This is illustrated in
Fig.~\ref{fig:2}.  Making this \ansatz\ in Eq.~\eqref{eq:rates},
and rearranging, shows that this point is reached on a time scale
\begin{equation}
\tcld\sim a_0^2/D_0\times\phi_0^{1/(\df-1)}({\lambda}/{a_0})^{2\df/(\df-1)}.
\label{eq:tcld}
\end{equation}
Once the clusters are stuck to the walls, no further changes occur, so
this is also the time to reach the final state of the system.  The
argument assumes $k_0\tcld\gg1$ but one can easily show that this is
true.  In fact, one can show that $\tcld\to\tppd$, and the aggregation
number of the characteristic clusters $N\to1$, as
$\phi_0\to(a_0/\lambda)^2$ from above, implying there is a smooth
crossover between the regime of primary particle deposition and the
present cluster deposition regime.

The time to reach the final state in Eq.~\eqref{eq:tcld} is an
increasing function of $\phi_0$ and at some point it must surpass the
gelation time in Eq.~\eqref{eq:tgel} which is a decreasing function of
$\phi_0$.  If this happens, the system will form a pore-space-filling
network \emph{before} it reaches the state where clusters become
attached to the walls.  To avoid this fate requires that
$\tcld\ll\tgel$, which after some rearrangement implies
$\phi_0\ll(a_0/\lambda)^{3-\df}$.  Thus the cluster deposition regime
occurs only in an intermediate range of volume fractions,
\begin{equation}
({a_0}/{\lambda})^2\ll\phi_0
\ll({a_0}/{\lambda})^{3-\df}.
\label{eq:cld}
\end{equation}
Note that this intermediate regime exists since $a_0\ll\lambda$, and
$\df>1$ is required for connected clusters.

Conversely, for
\begin{equation}
({a_0}/{\lambda})^{3-\df}\ll\phi_0,
\label{eq:gel}
\end{equation}
the final state is a pore-space-filling network or gel.  The time to
reach the final state in this situation is just the gelation time
given by Eq.~\eqref{eq:tgel}.  

At the crossover between cluster deposition and gelation, one has
$\tcld\sim\tgel\sim a_0^2/D_0\times({\lambda}/{a_0})^3$.  This is
found by setting $\phi_0\sim (a_0/\lambda)^{3-\df}$ in either of
Eqs.~\eqref{eq:tgel} or \eqref{eq:tcld}.

It is worth commenting on the properties of the clusters in cluster
deposition regime of Eq.~\eqref{eq:cld}.  After some algebra one finds
that the terminal characteristic cluster size, in units of the pore
size, obeys ${R}/{\lambda}\sim\phi^{1/2}$ where $\phi\sim nR^3$ is the
effective cluster volume fraction.  This shows that the clusters only
grow to span the pores as one approaches the crossover to the gelation
threshold where $\phi\to1$.  This, incidentally, is also the promised
\aposteriori\ justification for $R\ll\lambda$ mentioned above.  It
also eliminates a possible fate of the aggregating system: at no point
does the characteristic cluster size grow large enough to span the
pores ($R\sim\lambda$), without gelling ($\phi\ll1$).  Finally, by
analogy to the primary particle deposition case, the effective area
fraction covered by clusters in the cluster deposition regime is
$R^2n\lambda$.  But one has $R^2n\lambda\sim
R^3n(\lambda/R)\sim\phi/\phi^{1/2}\sim\phi^{1/2}$.  This means that
clusters deposit \emph{individually} in the cluster deposition regime,
and only start to touch each other as one approaches the gelation
threshold.

With these considerations it has now been demonstrated that the
crossovers between all the regimes are continuous.  At this point the
analysis is complete.  There are three predicted scaling regimes: a
primary particle deposition regime at the lowest primary particle
concentration, a cluster deposition regime at intermediate
concentrations, and a pore-space-filling network or gelation regime at
the highest concentrations.  This is shown schematically in
Fig.~\ref{fig:1}, using a slit to represent the pore space.  In
terms of the primary particle volume fraction, these
regimes are given by Eqs.~\eqref{eq:ppd}, \eqref{eq:cld} and
\eqref{eq:gel} respectively.  Likewise the time taken to reach the
final state in each of the regimes is given by Eqs.~\eqref{eq:tppd},
\eqref{eq:tcld} and \eqref{eq:tgel} respectively.  A regime of
pore-spanning clusters does not occur.

Fig.~\ref{fig:1} shows perhaps a surprising prediction, namely that
the time to reach the final state \emph{increases} with concentration
in the intermediate cluster deposition regime, reaching a maximum at
the gelation threshold.  Na{\"\i}vely one might have expected $\tppd$
to be an upper bound for the time to reach the final state.  After
all, non-aggregating particles would stick to the walls on a time
scale $\sim\tppd$ independent of the concentration.  However, in the
cluster deposition regime for an aggregating system, $\tppd$ is
initially much longer than the aggregation time.  What happens
then is that most of the primary particles form aggregates rather
than finding the walls.  The diffusion coefficient of an aggregate is
smaller than that of a primary particle, hence the time to reach the
final state (\ie\ the time to diffuse to the walls) must exceed
$\tppd$.

The present \letter\ is a self-contained presentation of a scaling
theory for DLCA in a porous medium.  Although the predictions seem
reasonable, and internally consistent, the approach is based at least
in part on the \ansatz, illustrated in Fig.~\ref{fig:2}, that there is
a crossover between cluster-cluster aggregation and cluster-wall
depletion.  For future work it is obviously prudent to test this.  An
obvious way is to use computer simulations.  For these, it is probably
sufficient to consider DLCA in a slit of width $\lambda$, as indicated
schematically in Fig.~\ref{fig:1}.  Experimental tests could also be
attempted although they are harder to make since a proper comparison
should ideally start from randomly dispersed primary particles which
are colloidally unstable.  Usually, colloidal instability is induced
by adding salt, or by some other mixing process, and in a porous
material it seems inevitable that this would involve flow fields.
However, ingenious methods have been devised for the
\insitu\ destabilisation of colloidal suspensions, for example using
the enzyme-catalysed hydrolysis of urea to increase the ionic strength
\cite{urea}.  Another possible way might be to shine UV light on
colloids which have been stabilised by photo-destructible surfactants
\cite{UVlight}, although so far as I know this latter method has not
been experimentally tested.

Two other directions for future work are as follows.  Firstly it would
be interesting to include the effects of a reduced sticking
probability, both for particle-particle collisions, and for
particle-wall collisions.  One might anticipate a variety of
cross-overs, based on our understanding of how reaction-limited
cluster aggregation (RLCA) crosses over to DLCA \cite{XWLH+85}.
Secondly for realistic applications it may be important to include the
effects of flow fields.  One can anticipate that flow would influence
the behaviour in two distinct ways.  Firstly, advection by the flow
may produce a different scaling behaviour for clusters formed at
Peclet numbers that are larger than unity \cite{DLCAsed}.  Secondly,
if the shear stresses are sufficiently large, they would have the
effect of rupturing clusters and therefore limiting growth
\cite{WB92}.

\begin{acknowledgments}

I have benefited from numerous discussions with my Unilever colleagues
Neil Shaw and Steve Wire; and with Robin Ball in the initial stages of
the work.  I also thank Ard Louis for drawing my attention to
Ref.~\cite{XBLC+08}.

\end{acknowledgments}


\begin{thebibliography}{19}
\expandafter\ifx\csname natexlab\endcsname\relax\def\natexlab#1{#1}\fi
\expandafter\ifx\csname bibnamefont\endcsname\relax
  \def\bibnamefont#1{#1}\fi
\expandafter\ifx\csname bibfnamefont\endcsname\relax
  \def\bibfnamefont#1{#1}\fi
\expandafter\ifx\csname citenamefont\endcsname\relax
  \def\citenamefont#1{#1}\fi
\expandafter\ifx\csname url\endcsname\relax
  \def\url#1{\texttt{#1}}\fi
\expandafter\ifx\csname urlprefix\endcsname\relax\def\urlprefix{URL }\fi
\providecommand{\bibinfo}[2]{#2}
\providecommand{\eprint}[2][]{\url{#2}}

\bibitem[{\citenamefont{Witten and Sander}(1981)}]{WS81}
\bibinfo{author}{\bibfnamefont{T.~A.} \bibnamefont{Witten}} \bibnamefont{and}
  \bibinfo{author}{\bibfnamefont{L.~M.} \bibnamefont{Sander}},
  \bibinfo{journal}{Phys. Rev. Lett.} \textbf{\bibinfo{volume}{47}},
  \bibinfo{pages}{1400} (\bibinfo{year}{1981}).

\bibitem[{DLC({\natexlab{a}})}]{DLCA83}
\bibinfo{note}{\bibinfo{author}{\bibfnamefont{P.}~\bibnamefont{Meakin}},
  \bibinfo{journal}{Phys. Rev. Lett.} \textbf{\bibinfo{volume}{51}},
  \bibinfo{pages}{1119} (\bibinfo{year}{1983});
  \bibinfo{author}{\bibfnamefont{M.}~\bibnamefont{Kolb}},
  \bibinfo{author}{\bibfnamefont{R.}~\bibnamefont{Botet}}, \bibnamefont{and}
  \bibinfo{author}{\bibfnamefont{R.}~\bibnamefont{Jullien}},
  \bibinfo{journal}{Phys. Rev. Lett.} \textbf{\bibinfo{volume}{51}},
  \bibinfo{pages}{1123} (\bibinfo{year}{1983}).}

\bibitem[{\citenamefont{Weitz and Oliveria}(1984)}]{WO84}
\bibinfo{author}{\bibfnamefont{D.~A.} \bibnamefont{Weitz}} \bibnamefont{and}
  \bibinfo{author}{\bibfnamefont{M.}~\bibnamefont{Oliveria}},
  \bibinfo{journal}{Phys. Rev. Lett.} \textbf{\bibinfo{volume}{52}},
  \bibinfo{pages}{1433} (\bibinfo{year}{1984}).

\bibitem[{DLA({\natexlab{a}})}]{DLA08}
\bibinfo{note}{\bibinfo{author}{\bibfnamefont{L.~A.} \bibnamefont{Turkevich}}
  \bibnamefont{and} \bibinfo{author}{\bibfnamefont{H.}~\bibnamefont{Scher}},
  \bibinfo{journal}{Phys. Rev. Lett.} \textbf{\bibinfo{volume}{55}},
  \bibinfo{pages}{1026} (\bibinfo{year}{1985});
  \bibinfo{author}{\bibfnamefont{T.~C.} \bibnamefont{Halsey}},
  \bibinfo{author}{\bibfnamefont{P.}~\bibnamefont{Meakin}}, \bibnamefont{and}
  \bibinfo{author}{\bibfnamefont{I.}~\bibnamefont{Procaccia}},
  \bibinfo{journal}{Phys. Rev. Lett.} \textbf{\bibinfo{volume}{56}},
  \bibinfo{pages}{854} (\bibinfo{year}{1986});
  \bibinfo{author}{\bibfnamefont{J.}~\bibnamefont{Lee}} \bibnamefont{and}
  \bibinfo{author}{\bibfnamefont{H.~E.} \bibnamefont{Stanley}},
  \bibinfo{journal}{Phys. Rev. Lett.} \textbf{\bibinfo{volume}{61}},
  \bibinfo{pages}{2945} (\bibinfo{year}{1988});
  \bibinfo{author}{\bibfnamefont{R.}~\bibnamefont{Blumenfeld}}
  \bibnamefont{and} \bibinfo{author}{\bibfnamefont{A.}~\bibnamefont{Aharony}},
  \bibinfo{journal}{Phys. Rev. Lett.} \textbf{\bibinfo{volume}{62}},
  \bibinfo{pages}{2977} (\bibinfo{year}{1989});
  \bibinfo{author}{\bibfnamefont{E.}~\bibnamefont{Somfai}},
  \bibinfo{author}{\bibfnamefont{L.~M.} \bibnamefont{Sander}},
  \bibnamefont{and} \bibinfo{author}{\bibfnamefont{R.~C.} \bibnamefont{Ball}},
  \bibinfo{journal}{Phys. Rev. Lett.} \textbf{\bibinfo{volume}{83}},
  \bibinfo{pages}{5523} (\bibinfo{year}{1999});
  \bibinfo{author}{\bibfnamefont{W.~G.} \bibnamefont{Hanan}} \bibnamefont{and}
  \bibinfo{author}{\bibfnamefont{D.~M.} \bibnamefont{Haffernan}},
  \bibinfo{journal}{Phys. Rev. E} \textbf{\bibinfo{volume}{77}},
  \bibinfo{pages}{011405} (\bibinfo{year}{2008}).}

\bibitem[{DLC({\natexlab{b}})}]{DLCAconc}
\bibinfo{note}{\bibinfo{author}{\bibfnamefont{M.}~\bibnamefont{Kolb}}
  \bibnamefont{and} \bibinfo{author}{\bibfnamefont{H.~J.}
  \bibnamefont{Herrmann}}, \bibinfo{journal}{J. Phys. A: Math. Gen}
  \textbf{\bibinfo{volume}{18}}, \bibinfo{pages}{L435} (\bibinfo{year}{1985});
  \bibinfo{author}{\bibfnamefont{M.~D.} \bibnamefont{Haw}},
  \bibinfo{author}{\bibfnamefont{W.~C.~K.} \bibnamefont{Poon}},
  \bibnamefont{and} \bibinfo{author}{\bibfnamefont{P.~N.} \bibnamefont{Pusey}},
  \bibinfo{journal}{Physica A} \textbf{\bibinfo{volume}{208}},
  \bibinfo{pages}{8} (\bibinfo{year}{1994}).}

\bibitem[{DLC({\natexlab{c}})}]{DLCAsed}
\bibinfo{note}{\bibinfo{author}{\bibfnamefont{A.~E.} \bibnamefont{Gonz\'alez}},
  \bibinfo{journal}{Phys. Rev. Lett.} \textbf{\bibinfo{volume}{86}},
  \bibinfo{pages}{1243} (\bibinfo{year}{2001});
  \bibinfo{author}{\bibfnamefont{C.~D.} \bibnamefont{Westbrook}} \etal,
  \bibinfo{journal}{Phys. Rev. E} \textbf{\bibinfo{volume}{70}},
  \bibinfo{pages}{021403} (\bibinfo{year}{2004});
  \bibinfo{author}{\bibfnamefont{A.~E.} \bibnamefont{Gonz\'alez}},
  \bibinfo{journal}{Europhys. Lett.} \textbf{\bibinfo{volume}{73}},
  \bibinfo{pages}{878} (\bibinfo{year}{2006}).}

\bibitem[{\citenamefont{Potanin and Uriev}(1990)}]{PU90}
\bibinfo{author}{\bibfnamefont{A.~A.} \bibnamefont{Potanin}} \bibnamefont{and}
  \bibinfo{author}{\bibfnamefont{N.~B.} \bibnamefont{Uriev}},
  \bibinfo{journal}{J. Coll. Int. Sci.} \textbf{\bibinfo{volume}{142}},
  \bibinfo{pages}{385} (\bibinfo{year}{1990}).

\bibitem[{\citenamefont{Wessel and Ball}(1992)}]{WB92}
\bibinfo{author}{\bibfnamefont{R.}~\bibnamefont{Wessel}} \bibnamefont{and}
  \bibinfo{author}{\bibfnamefont{R.~C.} \bibnamefont{Ball}},
  \bibinfo{journal}{Phys. Rev. A} \textbf{\bibinfo{volume}{46}},
  \bibinfo{pages}{R3008} (\bibinfo{year}{1992}).

\bibitem[{dee()}]{deepbed}
\bibinfo{note}{\bibinfo{author}{\bibfnamefont{M.}~\bibnamefont{Kolb}}
  \bibnamefont{and} \bibinfo{author}{\bibfnamefont{H.~J.}
  \bibnamefont{Herrmann}}, \bibinfo{journal}{J. Phys. A: Math. Gen}
  \textbf{\bibinfo{volume}{18}}, \bibinfo{pages}{L435} (\bibinfo{year}{1985});
  \bibinfo{author}{\bibfnamefont{M.~D.} \bibnamefont{Haw}},
  \bibinfo{author}{\bibfnamefont{W.~C.~K.} \bibnamefont{Poon}},
  \bibnamefont{and} \bibinfo{author}{\bibfnamefont{P.~N.} \bibnamefont{Pusey}},
  \bibinfo{journal}{Physica A} \textbf{\bibinfo{volume}{208}},
  \bibinfo{pages}{8} (\bibinfo{year}{1994}).}

\bibitem[{XBL()}]{XBLC+08}
\bibinfo{note}{\bibinfo{author}{\bibfnamefont{E.~S.} \bibnamefont{Boek}} \etal,
  \bibinfo{journal}{Energy Fuels} \textbf{\bibinfo{volume}{22}},
  \bibinfo{pages}{805} (\bibinfo{year}{2008}).}

\bibitem[{DLA({\natexlab{b}})}]{DLAwedge}
\bibinfo{note}{\bibinfo{author}{\bibfnamefont{T.~G.~M.} \bibnamefont{van~de
  Ven}}, \bibinfo{journal}{Coll. Surf. A} \textbf{\bibinfo{volume}{138}},
  \bibinfo{pages}{207} (\bibinfo{year}{1998});
  \bibinfo{author}{\bibfnamefont{N.~P.} \bibnamefont{Ryde}} \bibnamefont{and}
  \bibinfo{author}{\bibfnamefont{E.}~\bibnamefont{Matijevic}},
  \bibinfo{journal}{Coll. Surf. A} \textbf{\bibinfo{volume}{165}},
  \bibinfo{pages}{59} (\bibinfo{year}{2000}).}

\bibitem[{DLC({\natexlab{d}})}]{DLCAbooks}
\bibinfo{note}{See for example
  \bibinfo{editor}{\bibfnamefont{L.}~\bibnamefont{Pietronero}}
  \bibnamefont{and} \bibinfo{editor}{\bibfnamefont{E.}~\bibnamefont{Tosatti}},
  eds., \emph{\bibinfo{title}{Fractals in physics}}
  (\bibinfo{publisher}{North-Holland}, \bibinfo{address}{Amsterdam},
  \bibinfo{year}{1986});
  \bibinfo{author}{\bibfnamefont{R.}~\bibnamefont{Jullien}} \bibnamefont{and}
  \bibinfo{author}{\bibfnamefont{R.}~\bibnamefont{Botet}},
  \emph{\bibinfo{title}{Aggregation and fractal aggregates}}
  (\bibinfo{publisher}{World Scientific}, \bibinfo{address}{Singapore},
  \bibinfo{year}{1987});
  \bibinfo{author}{\bibfnamefont{T.}~\bibnamefont{Vicsek}},
  \emph{\bibinfo{title}{Fractal growth phenomena}} (\bibinfo{publisher}{World
  Scientific}, \bibinfo{address}{Singapore}, \bibinfo{year}{1989}).}

\bibitem[{\citenamefont{Meakin}(1984)}]{Mea84}
\bibinfo{author}{\bibfnamefont{P.}~\bibnamefont{Meakin}},
  \bibinfo{journal}{Phys. Rev. A} \textbf{\bibinfo{volume}{29}},
  \bibinfo{pages}{997} (\bibinfo{year}{1984}).

\bibitem[{Ley()}]{LeyvDE}
\bibinfo{note}{\bibinfo{author}{\bibfnamefont{F.}~\bibnamefont{Leyvraz}},
  \bibinfo{journal}{Phys. Rev. A} \textbf{\bibinfo{volume}{29}},
  \bibinfo{pages}{854} (\bibinfo{year}{1984});
  \bibinfo{author}{\bibfnamefont{P.~G.~J.} \bibnamefont{van Dongen}}
  \bibnamefont{and} \bibinfo{author}{\bibfnamefont{M.~H.} \bibnamefont{Ernst}},
  \bibinfo{journal}{Phys. Rev. Lett.} \textbf{\bibinfo{volume}{54}},
  \bibinfo{pages}{1396 } (\bibinfo{year}{1985}).}

\bibitem[{XWL()}]{XWLH+85}
\bibinfo{note}{\bibinfo{author}{\bibfnamefont{D.~A.} \bibnamefont{Weitz}}
  \etal, pp. \bibinfo{pages}{171--188} in
  \bibinfo{editor}{\bibfnamefont{R.}~\bibnamefont{Pynn}} \bibnamefont{and}
  \bibinfo{editor}{\bibfnamefont{A.}~\bibnamefont{Skjeltorp}}, eds.,
  \emph{\bibinfo{title}{Scaling phenomena in disordered systems}}
  (\bibinfo{publisher}{Plenum}, \bibinfo{address}{New York},
  \bibinfo{year}{1985}).}

\bibitem[{\citenamefont{Meakin}(1992)}]{Meakin92}
\bibinfo{author}{\bibfnamefont{P.}~\bibnamefont{Meakin}},
  \bibinfo{journal}{Physica Scripta} \textbf{\bibinfo{volume}{46}},
  \bibinfo{pages}{295} (\bibinfo{year}{1992}).

\bibitem[{\citenamefont{von Smoluchowski}(1916)}]{Smol16}
\bibinfo{author}{\bibfnamefont{M.}~\bibnamefont{von Smoluchowski}},
  \bibinfo{journal}{Phys. Z.} \textbf{\bibinfo{volume}{17}},
  \bibinfo{pages}{593} (\bibinfo{year}{1916}).

\bibitem[{ure()}]{urea}
\bibinfo{note}{\bibinfo{author}{\bibfnamefont{L.~J.} \bibnamefont{Gauckler}},
  \bibinfo{author}{\bibfnamefont{T.}~\bibnamefont{Graule}}, \bibnamefont{and}
  \bibinfo{author}{\bibfnamefont{F.}~\bibnamefont{Baader}},
  \bibinfo{journal}{Mater. Chem. Phys.} \textbf{\bibinfo{volume}{61}},
  \bibinfo{pages}{78} (\bibinfo{year}{1999});
  \bibinfo{author}{\bibfnamefont{S.}~\bibnamefont{Romer}},
  \bibinfo{author}{\bibfnamefont{F.}~\bibnamefont{Scheffold}},
  \bibnamefont{and}
  \bibinfo{author}{\bibfnamefont{P.}~\bibnamefont{Schurtenberger}},
  \bibinfo{journal}{Phys. Rev. Lett.} \textbf{\bibinfo{volume}{85}},
  \bibinfo{pages}{4980} (\bibinfo{year}{2000}).}

\bibitem[{UVl()}]{UVlight}
\bibinfo{note}{\bibinfo{author}{\bibfnamefont{I.~R.} \bibnamefont{Dunkin}}
  \etal, \bibinfo{journal}{J. Chem. Soc., Perkin Trans.}
  \textbf{\bibinfo{volume}{2}}, \bibinfo{pages}{1837} (\bibinfo{year}{1996}).}

\end{thebibliography}

\end{document}